\documentclass[aip, preprint]{revtex4-1}
\usepackage{amsmath}    
\usepackage{graphicx}   
\usepackage{textcomp}

\usepackage{hyperref, color}
\definecolor{linkColor}{rgb}{0.8,0,0}
\definecolor{darkred}{rgb}{0.8,0,0}
\hypersetup{pdfborder={0 0 0},colorlinks=true,urlcolor=linkColor,citecolor=linkColor}

\begin{document}

\title{Design of a Graphical User Interface for Few-Shot Machine Learning Classification of Electron Microscopy Data}

\author{Christina Doty*}
\affiliation{Department of Materials Science and Engineering, University of Washington, Seattle, Washington 98195}

\author{Shaun Gallagher*}
\affiliation{Department of Chemistry, University of Washington, Seattle, Washington 98195}

\author{Wenqi Cui*}
\affiliation{Department of Electrical and Computer Engineering, University of Washington, Seattle, Washington 98195}

\author{Wenya Chen*}
\affiliation{Department of Chemical Engineering, University of Washington, Seattle, Washington 98195}

\author{Shweta Bhushan*}
\affiliation{Department of Materials Science and Engineering, University of Washington, Seattle, Washington 98195}

\author{Marjolein Oostrom}
\affiliation{National Security Directorate, Pacific Northwest National Laboratory, Richland, Washington 99352}

\author{Sarah Akers}
\affiliation{National Security Directorate, Pacific Northwest National Laboratory, Richland, Washington 99352}

\author{Steven R. Spurgeon}
\thanks{*These authors contributed equally}
\email{steven.spurgeon@pnnl.gov}
\affiliation{Energy and Environment Directorate, Pacific Northwest National Laboratory, Richland, Washington 99352}

\date{\today}

\begin{abstract}
The recent growth in data volumes produced by modern electron microscopes requires rapid, scalable, and flexible approaches to image segmentation and analysis. Few-shot machine learning, which can richly classify images from a handful of user-provided examples, is a promising route to high-throughput analysis. However, current command-line implementations of such approaches can be slow and unintuitive to use, lacking the real-time feedback necessary to perform effective classification. Here we report on the development of a Python-based graphical user interface that enables end users to easily conduct and visualize the output of few-shot learning models. This interface is lightweight and can be hosted locally or on the web, providing the opportunity to reproducibly conduct, share, and crowd-source few-shot analyses.
\end{abstract}

\maketitle

\section{Introduction}

High-resolution characterization of materials using electron microscopy is an essential part of advancing technology in various fields, such as clean energy, catalysis, biomedicine, and quantum computing.\cite{Rai2009,Zhang2019,Pennycook2017,Shah2019,Pennycook2006} The techniques of scanning electron microscopy (SEM) and (scanning) transmission electron microscopy ((S)TEM) have provided fundamental new insight into materials structure and chemistry, with the latter capable of routine imaging and spectroscopy at the atomic scale.\cite{MacLaren2014} (S)TEM in particular is widely used to study phase distributions, microstructures, and crystallographic defects in nanomaterials to build more accurate structure-property models.\cite{Fullwood2010} Traditional methods of image analysis involve an expert user manually observing and analyzing individual images for specific microstructural features; not only is this approach tedious, it can also be error-prone, is often not reproducible, and is difficult to scale to large volumes of data.\cite{Tropsha2017,Plant2014} With the introduction of high-speed detectors and spectrometers,\cite{Tate2016a,Mir2017,Plotkin-Swing2020,MacLaren2020} today's instruments now generate data at scales far beyond the range of direct human comprehension (up to $\sim200$ Tb/hr), which has spurred efforts to integrate high-throughput data science tools into the microscopy workflow.\cite{Spurgeon2020c,Ede2020,Kalinin2019,Belianinov2015}

One of the most common and useful tasks in electron microscopy is that of semantic segmentation and phase identification.\cite{DeCost2019} In this task, the abundance and spatial location of relevant microstructural features or motifs is classified, with the goal to derive physically meaningful descriptors for a dataset.\cite{Vlcek2017,Vasudevan2016} While small datasets may be amenable to manual analysis, emerging automated systems\cite{Schorb2019} require approaches capable of analyzing tens to hundreds of thousands of images containing dense, noisy, or rare features of interest at high speed. Moreover, these approaches must be scalable to multiple imaging and spectroscopic modalities to harness the full potential of modern instrumentation and derive more unique structure-property solutions.\cite{Spurgeon2020c} To address this challenge, a range of commercial\cite{MIPAR,TFAvizo} and open-source\cite{CellProfiler,ImageJ,SerialEM, Schorb2019} packages have been developed for full and semi-automated image analysis, particularly in the biological sciences. While each has its strengths and weaknesses, these packages are often require extensive manual tuning, particularly in cases where limited prior labeled data is available.

Over the past several years, machine learning (ML)-based approaches have been developed for segmentation, classification, and other tasks in electron microscopy.\cite{Ede2020,Aguiar2019a,Kannan2018,Vasudevan2018, Ziatdinov2017a, Voyles2016, DeCost2015} Put simply, these methods train a neural network to learn the mapping between the latent space representation of image features and their microstructural classification. However, most ML frameworks require a large support set of labeled images for training (e.g., containing $>100$ examples). Since labeling can require time-consuming manual identification and annotation of image features,\cite{Iren2021} preparing large training sets is typically not practical, especially when considering that feature types, imaging artifacts, and data quality can vary greatly from experiment to experiment.\cite{Hattar2021,Taheri2016} 

Few-shot ML has been proposed as a solution to the problem of limited training data.\cite{Hilliard2018,Finn2017} In this approach, a small, expert-selected training dataset is examined by a meta-learner that extracts a latent space prototype vector for each expert-provided label. These labels are based on the expert's \textit {a priori} knowledge of the image or past ground truth examples. Classification is then performed by computing Euclidian distances to the prototype representations. Few-shot learning of TEM images has recently been successfully demonstrated for a variety of microstructures\cite{Akers} and has been shown to efficiently generalize to different labeling tasks. However, the actual usage of the few-shot code requires multiple steps of image processing, support set selection, and few-shot classification. This process can be difficult for those not familiar with ML and programming, posing a barrier to real-world use.

Here we describe the design and implementation of a graphical user interface (GUI) for few-shot ML-based segmentation of electron microscopy data. We present an intuitive Python Flask-based application (pyChip) to upload image data, select desired features, perform classification, and visualize statistics on feature distributions. The user-friendly interface provides dynamic, real-time feedback that allows the expert to tune their analysis for different microstructural features, classification tasks, length scales, and image types. While we present TEM example data, we emphasize that this application can be extended to other types of imaging, such as SEM. In addition, the application is portable to both local and web-based hosts and can be easily updated with different back-end codes for few-shot learning. It is also lightweight and can be run on an average laptop computer, which will facilitate its adoption by the microscopy community.

\section{Results}

\begin{figure}
\includegraphics[width=\textwidth]{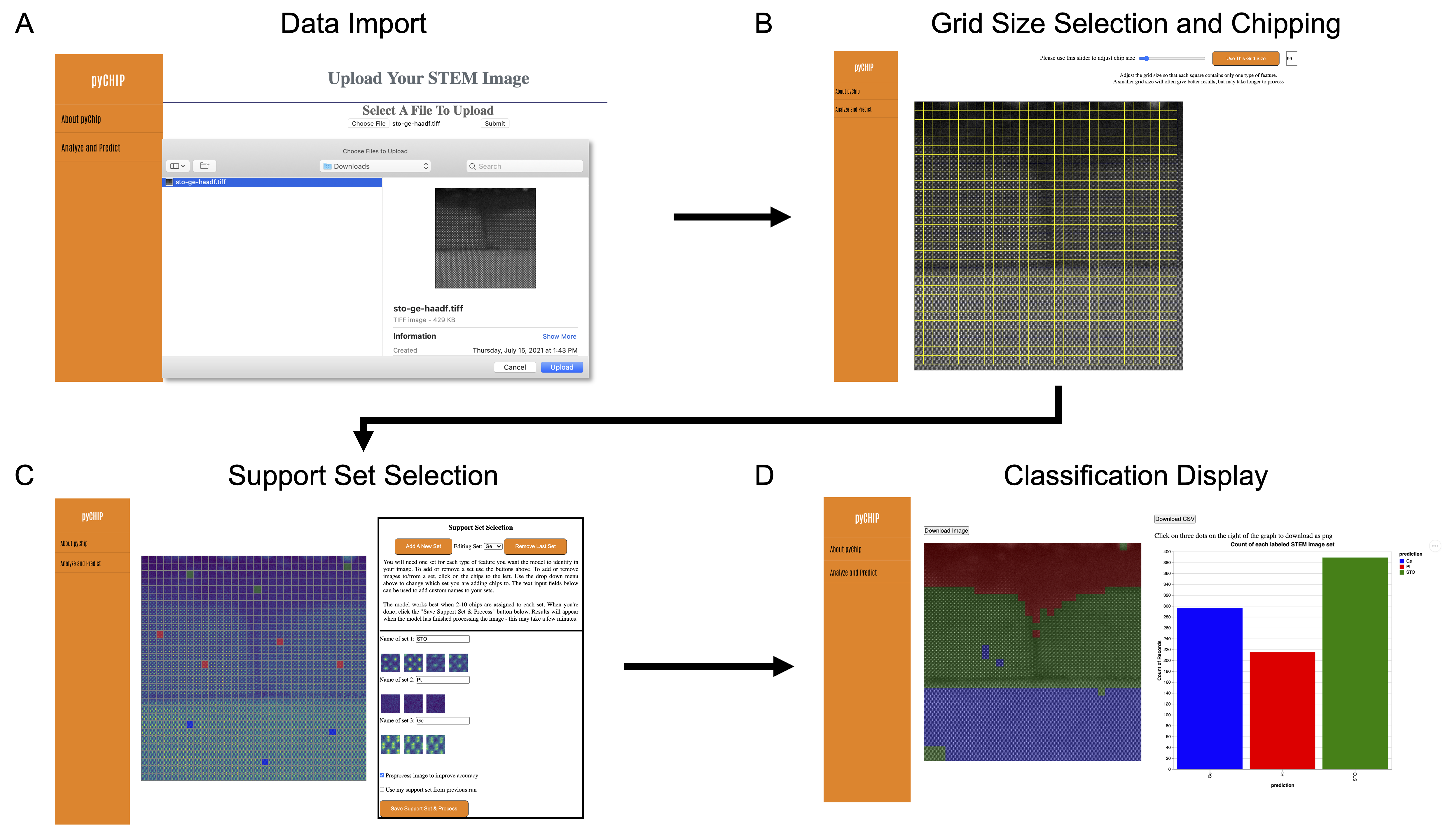}
\caption{(A--D) Overview of the pyChip application workflow for few-shot classification of electron microscopy data, consisting of data import, grid size selection and chipping, support set selection, and classification display, respectively. \label{fig:UI-overview}}
\end{figure}

An overview of the pyChip application workflow is shown in Figure \ref{fig:UI-overview}. The pyCHIP user interface contains a homepage and components for image data import, grid selection, support set selection, feature identification, and classification display. Each component is assigned a respective HTML page, serviced by a unique route. The tab on the left is used for navigation. The following sections describe the steps in the analysis workflow and practical considerations for their use. These sections are followed by selected examples of the analysis of different material microstructures. Finally, we conclude with an outlook on potential future developments and the application's usage in emerging automated analysis workflows.

\subsection{Data Import}

As shown in Figure \ref{fig:UI-overview}.A, the pyCHIP application initially presents the user with an upload prompt to select images for analysis. Currently allowed data formats are \texttt{jpg}, \texttt{png}, \texttt{tiff}, and \texttt{dm4}. The first three are common image formats, while the last is a proprietary container format used by the Gatan Microscopy Suite,\cite{Gatan} an industry-standard TEM analysis application. Images should be imported without any visible scale bar or other annotation. The \texttt{dm4} format is loaded using a library developed by the National Center for Electron Microscopy\cite{NCEM} and pixel values are created by scaling the embedded image data to a range between 0 and 255. If no image is uploaded before submitting, an error message indicates that no appropriate file was selected. While in principle any image resolution can be used, the processing time does increase significantly with larger resolution. For best performance, it is recommended to downsample large images to resolution of $512 \times 512$ or $1024 \times 1024$ pixels. However, it is important to avoid aliasing artifacts, which can occur in high-resolution micrographs, particularly those containing lattice features.

\subsection{Grid Size Selection and Chipping}

\begin{figure}
\includegraphics[width=\textwidth]{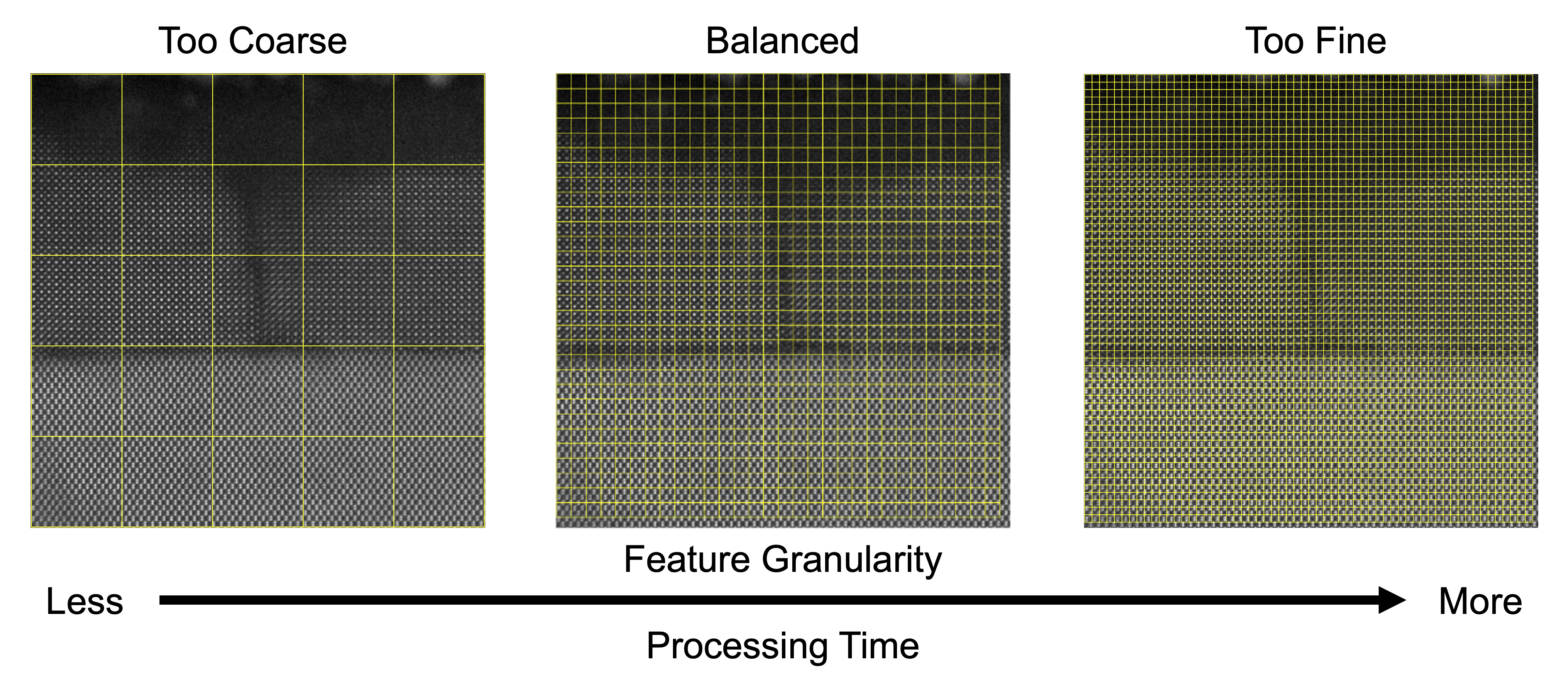}
\caption{Grid size selection relative to atomic-scale motifs for effective support set selection, illustrating trends in processing time vs. feature granularity. \label{fig:UI-grid}}
\end{figure}

After the selected data has been imported, the user must perform the important task of grid size selection and image chipping, as shown in Figure \ref{fig:UI-overview}.B. The user selects an appropriate sampling grid relative to their features of interest, which is used to chip the image in the backend few-shot code. To aid in this process, the user can drag a slider bar to dynamically adjust a superimposed grid atop their image. An ideal support set for the few-shot pyCHIP model is built from a grid with each square containing a unique microstructural feature or motif. The range of allowed grid sizes is dynamic, scaling from a minimum 50 pixels up to the dimensions of the uploaded image. The default lower limit is set to 50 pixels for visibility, but can be as small as the single pixel level. Too coarse a grid size will not capture relevant features, while too fine a grid will greatly increase processing time, so the user should test different sizes to strike an optimal balance, as shown in Figure \ref{fig:UI-grid}. Once an appropriate size is determined, the user clicks the ``Use This Grid Size'' button and the desired grid size is passed to a function that crops the edges of the full size image to the total mesh, sub-divides the image, and saves the resulting collection of support set ``chips.'' Both the raw chips and support sets are saved for future recall.

\subsection{Support Set Selection}

After chipping, the user must assign support set examples as features of interest. Building on the concept of the ubiquitous CAPTCHA (Completely Automated Public Turing test to tell Computers and Humans Apart) security feature\cite{VonAhn2004} on websites, we have designed an intuitive interface for labeling. The chips created in the previous step now appear on the dashboard as interactive objects. As shown in Figure \ref{fig:UI-overview}.C, the user is provided with an empty starting support set in a drop-down menu, which can be renamed. This menu controls the specific support set(s) that is/are available for modification. Any image chips that the user clicks on will be added to the currently staged support set and appear in the right panel. These selected chips will appear highlighted in colors based on the user-designated support set. In cases where the user clicks the wrong chip and would like to remove it from the support set, they may click the highlighted chip again. Additional support sets can be added or removed by clicking the appropriate buttons at the top of the panel.

In practice, the user is is advised to assign a minimum of 2 chips or maximum of 10 chips to each labeled set.\cite{Akers} These chips should be selected from different parts of the image to capture some of the variability in the data. Since the selected chips will become the support set for the few-shot ML model, all suspected feature types should be selected---sets can easily be adjusted by reprocessing. After the chip selection and labeling, the user may click ``Save Support Set \& Process'' to update the training set and begin training the model. The trained model will then predict features against the selected support sets.

\begin{figure}
\includegraphics[width=\textwidth]{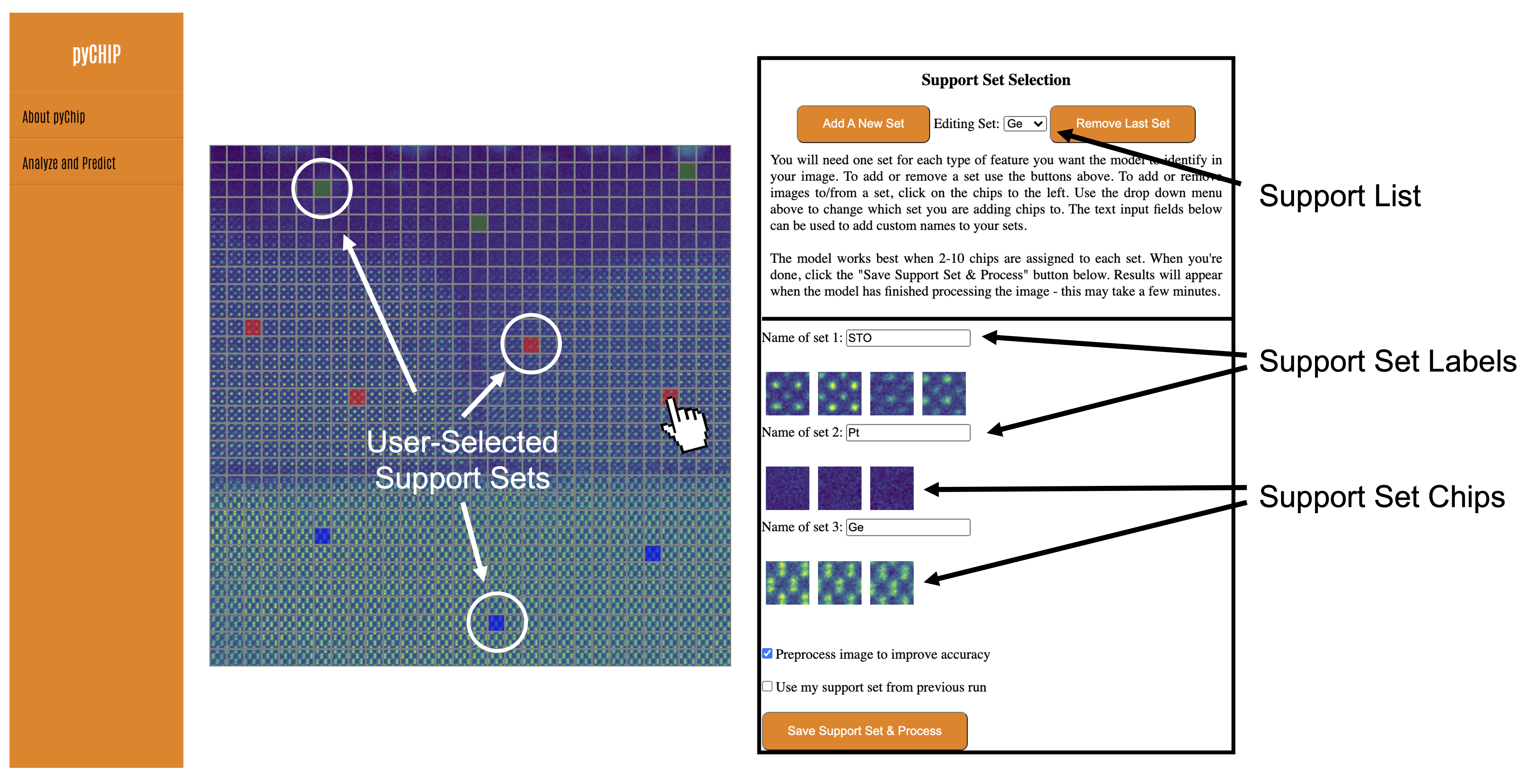}
\caption{Interactive support set selection and labeling of user-defined features of interest. \label{fig:UI-chipping}}
\end{figure}

\subsection{Classification Display}

After training the model using the selected chips, the model predicts the support set label for all the chips in the original image and returns a colorized segmented image, as shown in the left of Figure \ref{fig:UI-overview}.D. The right side presents the count for each user-selected support set within the image. The segmented image and associated statistics tell the user critical information regarding features in the image. The user can export the colorized image (\texttt{png} format), the bar chart (\texttt{png} or \texttt{svg} formats), and the raw data for the bar chart (\texttt{csv} format).

\subsection{Usage Examples}

\begin{figure}
\includegraphics[width=\textwidth]{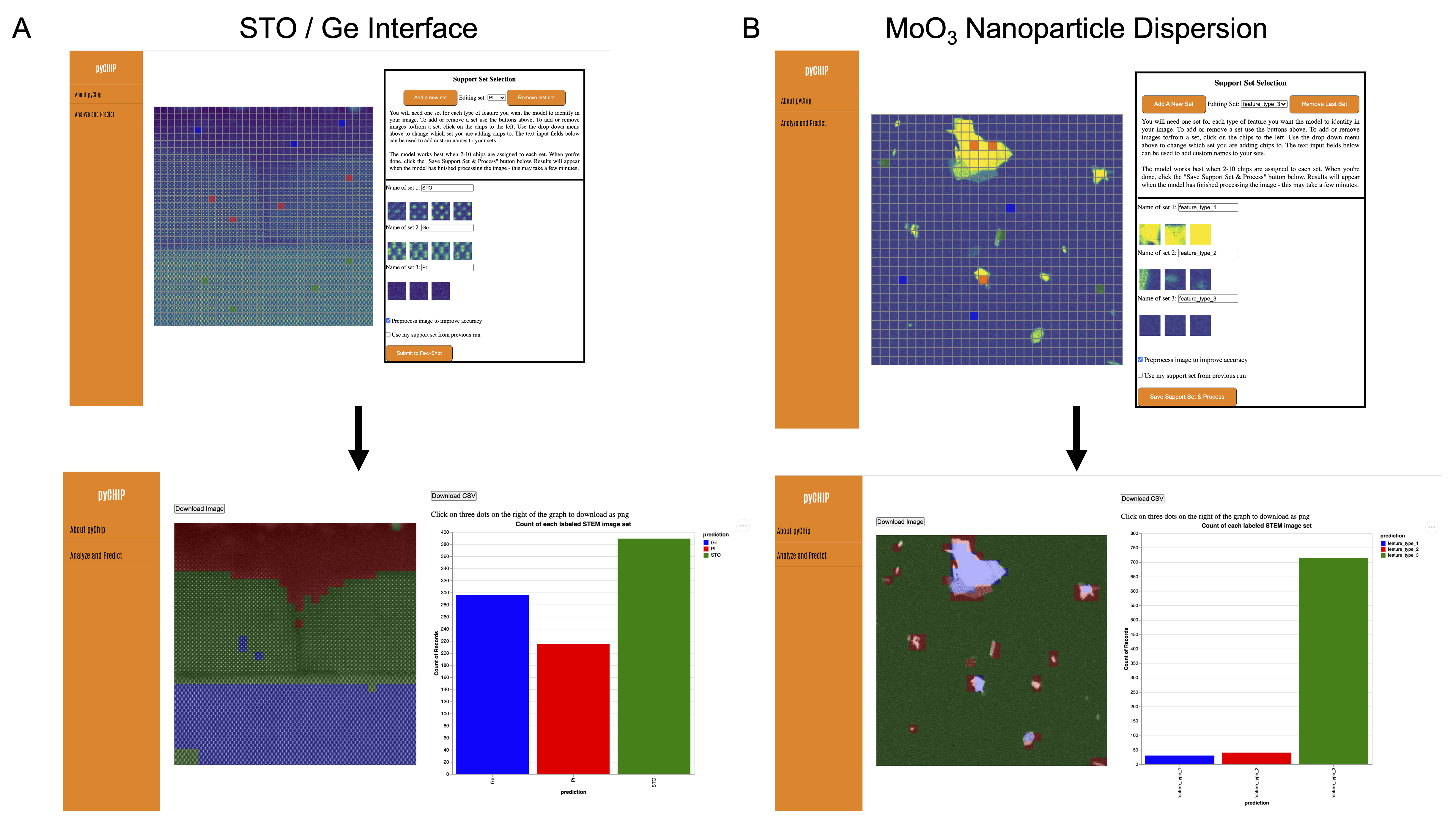}
\caption{(A--B) Support set selection and output visualization for a cross-sectional STO / Ge interface and a dispersion of MoO$_3$ nanoparticles on a carbon support, respectively. \label{fig:Usecases}}
\end{figure}

Here we discuss two specific material use cases to illustrate the process of support set selection and corresponding outputs. Nanomaterials are a natural subject for study via TEM, since they exhibit properties distinct from bulk materials that are strongly governed by atomic-to-nanoscale structure and defects. Quantifying the distribution of these features, such as the abundance of particular phases, defects, and morphology, is a common but challenging and time-consuming task.\cite{Pazdernik2020,Ede2020} We consider two very different examples: a cross-sectional thin film heterostructure of SrTiO$_3$ (STO) / Ge and nanoparticles of MoO$_3$. The former sample represents a potential material for next-generation eleectronics\cite{Chambers2017,Hudait2015} and contains features, such as interfaces and crystal motifs, that are commonly encountered in atomic-scale images ($>6$ Mx magnification). The latter sample is an important precursor for organic photovoltaic synthesis\cite{Gong2020} and takes the form of a particle dispersion with varying morphology that is evident at low magnification ($30-80$ kx). This example is particularly relevant, since many important systems manifest in nanoparticle form. As shown Figure \ref{fig:Usecases}.A, the atomic-resolution STO / Ge image is divided into square chips, each 99 pixels wide relative to a total image width of 2970 pixels. By selecting just three or fewer examples from each region of interest (STO, Ge, and vacuum), the user is able to quickly perform a segmentation that traces the boundaries of the interface and the sample surface. Moreover, the user can generate statistics on the abundance (pixel fraction) of each feature type. In the case of the lower-magnification MoO$_3$ nanoparticle sample, shown in Figure \ref{fig:Usecases}.B, the image is divided into square chips, each 18 pixels wide relative to a total image width of 504 pixels. The user provides three examples each of two different particle morphologies and the carbon background, to quickly distinguish features and produce statistics. It is important to note that both the labeling task and image type can be easily changed \textit{without} time-consuming retraining of the network, which is a strength of the few-shot approach.\cite{Akers} The pyCHIP application provides rapid feedback, so the user can adjust the grid size and select different support sets if necessary. This approach enables both single image analysis and more systematic comparisons of images of similar type.

\section{Discussion}

We have described the design and implementation of a flexible, multi-platform GUI application for data import, support set selection, classification, and visualization of a few-shot ML code. This interface will aid in development of rapid, reproducible, and scalable analyses of electron microscopy images. We envision that this tool will eventually become part of a standard workflow for data classification, particularly in cases where there are limited prior labeled examples.

Moving forward, it will be desirable to implement both usability and model improvements. These include the implementation of user accounts and recall of images, parameters, and support sets. In addition, more automatic guidance on chipping and support set selection is necessary, as well as recommendations for appropriate parameters. Although the colorized segmentation and bar graph provide a clear summary of the location and abundance of each feature, future developments should include more informative statistics such as error bars or other confidence measures. While this task alone is a significant challenge and ongoing area of research in ML, such information would provide the user with more complete, insightful semantics for their analysis.

Finally, it will be important to expand the application to provide detailed management of metadata and curation of the labeling process, particularly for multimodal datasets.\cite{Withers2019} These procedures can eventually be automated, enabling more repeatable image analysis and leading to improved reproducibility in materials characterization.\cite{Tropsha2017} We envision on-the-fly integration of the few-shot approach to conduct analyses and track the dynamic evolution of systems in real-time, which will be an important step toward autonomous microscopy.

\section{Methods}

The application integrates Python, D3, JavaScript, HTML/CSS, and Vega-lite with Flask, a Python web framework. Flask allows the back-end python scripts to easily pass information to the front-end interactive visualization.\cite{grinberg2018flask}

The front-end interactive visualization was created with JavaScript and HTML/CSS. JavaScript can manipulate Document Object Model (DOM) objects such as paragraphs and images.\cite{flanagan1998java} JavaScript was used to allow users to (1) view the image chips, (2) modify the names of support sets, (3) assign image chips to a specified support set, and (4) view the current collection of selected chips for each support set. Parts of the JavaScript code utilized D3, an open-source JavaScript library capable of building customized visualization from code.\cite{bostock2011d3} D3 uses data as input to create visualizations with Scalable Vector Graphics (SVG) elements such as lines, rectangles and circles. In this application, D3 was used to dynamically display grid lines over the user's image based on a user-inputted grid-size. The code for the grid lines was adapted from code written by user "Chuck Grimmett" on \href{https://bl.ocks.org/cagrimmett/07f8c8daea00946b9e704e3efcbd5739}{bl.ock}. \cite{d3grid} We also used Vega-lite, a JSON syntax to display the results of the few-shot model in the form of a bar chart.\cite{satyanarayan2016vega}

The back-end scripts are written in Python. The first function chips the images into squares matching the user-selected grid size. The second function implements the few-shot ML model with the selected support sets as input. The few-shot function utilizes a pre-trained ResNet available from PyTorch.

The Flask Framework allows the inputs from the front-end user interaction to be passed as input to the Python scripts on the back-end. The user selected grid-size on the front-end is used as an input for the image chipping Python function on the back-end. Likewise, the support set selected on the front-end is used as an input for the few-shot Python functions on the back-end. In turn, outputs from the back-end Python scripts, such as the image chips and the few-shot model's results are displayed in the front-end web page. The Flask framework mediates these front-end and back-end interactions.

\clearpage

\section{Data Availability}

The raw images shown are available on FigShare at \url{https://doi.org/10.6084/m9.figshare.14850102.v1}

\section{Code Availability}

The GUI code used in this study is available on Github at \url{https://github.com/pnnl/pychip_gui}. The core few-shot codebase is proprietary, but the Prototypical Network code is available on Github at \url{https://github.com/jakesnell/prototypical-networks}.

\section{Acknowledgements}

The authors would like to thank Drs. Jenna Pope, Elizabeth Kautz, and Matthew Olszta for reviewing the manuscript. This research was supported by the I3T Commercialization Laboratory Directed Research and Development (LDRD) program at Pacific Northwest National Laboratory (PNNL). PNNL is a multiprogram national laboratory operated for the U.S. Department of Energy (DOE) by Battelle Memorial Institute under Contract No. DE-AC05-76RL0-1830. Experimental sample preparation was performed at the Environmental Molecular Sciences Laboratory (EMSL), a national scientific user facility sponsored by the Department of Energy's Office of Biological and Environmental Research and located at PNNL. TEM data was collected in the Radiological Microscopy Suite (RMS), located in the Radiochemical Processing Laboratory (RPL) at PNNL. C.D.,S.G., W.C., W.Q. and S.B. acknowledge support from the University of Washington Clean Energy Institute and the National Science Foundation Research Traineeship under Award NSF DGE-1633216.

\section{Author Information}

S.A., M.O., and S.R.S. conceived and developed the project plan. C.D., S.G., W.C., W.C., S.B., M.O., and S.A. designed the user interface and developed the code for its implementation. All authors contributed to the writing and editing of the manuscript.

\section{Competing Interests Statement}

The authors declare no competing interests.

\clearpage

\bibliography{references, oostrom_refs}
\bibliographystyle{plainurl}

\end{document}